# ATTACKS AND COUNTERATTACKS ON WIRELESS SENSOR NETWORKS

Nusrat Fatema[1] and Remus Brad[2]

[1]Faculty of Engineering, B-dul Victoriei 10,
Lucian Blaga University of Sibiu, Sibiu, Romania

[2]Computer Science and Electrical Engineering Department, B-dul Victoriei 10,
Lucian Blaga University of Sibiu, Sibiu, Romania

## ABSTRACT

*WSN is formed by autonomous nodes with partial memory, communication range, power, and bandwidth. Their occupation depends on inspecting corporal and environmental conditions and communing through a system and performing data processing. The application field is vast, comprising military, ecology, healthcare, home or commercial and require a highly secured communication. The paper analyses different types of attacks and counterattacks and provides solutions for the WSN threats.*

## 1. INTRODUCTION

WSN is a technology for data processing which has tiny sensor nodes, operating autonomously, characterized by strictly guarded computational, power resources and ad hoc environment. It can be used for many different applications range armed implementations in the battleground, environmental monitoring, in health sectors as well as emergency responses and various surveillances. Wireless sensor networks require security mechanisms, as it may interrelate with sensitive data and function in antagonistic environments. Due to different constraints of resource and computation, security in sensor networks poses numerous challenges. In actuality, it is more susceptible to a variety of security intimidation as the freehanded communication medium is more vulnerable to security attacks than those of the guided communication medium.

Security mechanism is necessary for all types of WSN to guarantee the functionality of WSN in wicked environment. It is a great challenge due to the restriction of resources in the sensor nodes of a WSN.

The following document has: Section 2 presents a synopsis of the security requirements for WSN, Section 3 has attacks/ threats based on potentiality of the attacker, Section 4 consists of the attacks/ threats based on the location of the attacker, and Section 5 covers the attacks based on protocol layer followed by its counter attacks. Section 6 has different types of attacks are compared and followed by conclusions.

## 2. SECURITY REQUIREMENTS

The security requirements of a WSN are data integrity, confidentiality and freshness, time synchronization, authentication, availability, self organization and secure localization according to [1].

Data integrity ensures that the information do not alter in transfer, either due to accident or by malicious intent, as applications for example surveillance based on privacy. Keeping confidentiality is done by encryption. Even if data integrity and confidentiality are met we still need to send the fresh messages.

In data freshness, the data should be recent and with no message replayed. To ensure it, a time stamp is added to each packet. And also the receiver has to make sure that the data used in decision-making process originates from the correct source as WSN sends sensitive data.

To exchange messages authentication is always mandatory. Sensor nodes are powered by batteries and this may run out due to surplus calculation or transmit. This happens when attacker jam the link to make sensors busy.

According to [1], in order to maintain the availability of the network and also to avoid harming the operation of the network, security is highly noteworthy. WSN considers every sensor node is self-sufficient and supple enough to organize and heal them. Every node does random operations so particularly for WSN management no fixed infrastructure is there. WSN should organize them to maintain multi-hop routing, carry out key supervision and build conviction among sensors. Some applications on sensor network depend on time management. Radio of the sensor should be turned off at times to save power. Sensor network always needs accurate information of the position so it is easy for an attacker to manipulate sensors to a non-secured location just by weakening the signal strengths.

## 3. ATTACKS BASED ON POTENTIALITY OF THE ATTACKER

According to [2], there are four types of security threats:

- **Interception**: It happens when an illicit party gain right of entry. Attacker is always ready to gain illegal access to sensor node. This is a risk to secrecy/confidentiality. Example: wiretapping for capturing information from the network.

- **Interruption**: It happens when a feature is damaged or not available. This attack is on the accessibility of the system. Example: disconnecting the communication, physical capture of nodes, insertion of wicked code and corruption of information.

- **Modification**: It happens when an unofficial party gains access and mess with the asset. It is a threat to integrity. Example: by amending a value in the packets being broadcast / causing a DoS attack like down pouring the network with sham.

- **Fabrication**: It's a threat to integrity and authentication. It happens when an illicit bash pops in forge object into the system. Here an attacker adds wrong data and negotiate the fidelity of the record. Example: adding up of data to a file.

WSNs are endangered to security attacks because of the broadcast feature of the communication means and as well as to the assignment of nodes in an unfriendly or dangerous environment, physically unprotected. Main security attacks based on capability of the network are of two types: passive and active attack.

## 3.1. Passive Attacks

The attacker indulges in eaves dropping, monitoring of data transmission. It endeavours to learn or to utilize the information from the method but doesn't affect the resources or do not intend to damage other nodes. The attacker aims to obtain information that is in transit, as the term passive indicates that the attacker does not attempt to perform any modifications to the data. The following methods can be used:

- **Traffic Analysis**: To facilitate an adversary just to harm the sensor network or to analyse the communication patterns, sensor activities needs to reveal adequate information [3].

- **Eavesdropping and Monitoring**: Snooping to the data the foe could easily learn the communication data. Eavesdropping can act successfully against the privacy shield. The control packets which contain more thorough information are reachable through the location server [3].

- **Attacks against Privacy**: Through direct site observation much data from networks could probably be collected. So the network intensifies the privacy quandary since they manage much information available easily through distant access. Hence, foes need not be present physically to uphold observation [4].

- **Camouflage Adversaries**: Here a spiteful node can be added as a normal node and then these nodes pretend to be a normal node to create an attention to the packets and then misroute the packets and thus accomplishing the privacy analysis [3].

## 3.2. Active Attacks

In this attack spiteful nodes can harm other nodes by creating outage of the network. It involves a few alterations of the data or the formation of a fake stream. These attacks cannot be prevented easily. Attacks can be sub divided into 4 types:

- **Masquerade**: This happens when one unit pretends to be dissimilar unit. Example: After valid verification sequence has taken place these sequences can be replaced or captured thus allowing an endorsed unit with few privileges to gain extra privilege by mimicking a body that has those privileges. Interruption attacks can be called masquerade attacks.

- **Replay**: It engages the unreceptive capture of a data unit and its following retransmission to fabricate an illicit effect.

- **Modification of Message**: It implies some change to the original message and produces an unauthorized effect.

- **Denial of Services**: Caused by fabrication, it thwarts the normal use and supervision of facilities for the communication of data. Denial disables the network, produces disturbance of the total network or overloads it with unnecessary data in order to mortify the performance. It is difficult to prevent active attack due to the extensive assortment of physical and network threats. Here a foe effort to stop authorized client in using services of the network. This attack not only meant for the foes to disrupt/ destroy a network but also lessen the capability of the network to offer a good service. In different networking layers these types of attacks might be performed [4].

# 4. ATTACKS BASED ON THE LOCATION OF THE ATTACKER

According to [5], attackers can be deployed inside (internal attack) or outside (external attack). Based on attackers' location this section classifies the WSNs' link layer attacks.

## 4.1. External Attacker (outsider)

This attack can be defined as a physical one, where the attacker doesn't have any information about the cryptographic internal information of the network. Some of the most common features are:

- Mote/Laptop class devices
- Committed by illegitimate parties
- External to the network (out of WSN range)
- Commencing attacks without even being authenticated

Some other effects of these attacks:

- Consumption of resources in WSNs
- Jamming the entire communication
- Triggering DoS attacks

## 4.2. Internal Attacker (insider)

When a valid node of the network act unusually it is considered as an internal attack. It utilizes the compromised node to attack the network and can demolish or interrupt the network easily. It is one of the foremost disputes in WSNs, having foundation from the inner WSNs and accessing all other nodes inside its limit.

Important goals of these types are:

- Entrance to WSN codes
- Entrance to cryptography keys
- Threat: to the efficiency of the network
- Partial/total disruption
- Revealing secret keys.

# 5. ATTACKS BASED ON THE PROTOCOL LAYER AND ITS COUNTERATTACK

Based on protocol layer the attacks on WSN are described below.

## 5.1. Sink Hole

**Attack:**

- Scheme: By advertising a foe generates metaphorical sinkhole. Example: high quality way to support station.
- Attacker like laptop class can provide this kind of course linking all nodes to genuine sink and after that drop packets selectively.

- Almost all transfer is heading to the counterfeit sinkhole.
- WSNs are highly vulnerable to this attack because of the pattern of the communication: Lots of transfer is heading towards sink as like single point of failure.

**Counter Attacks:**

- Use of a typical system which will alert every node so that a pertinent node will not listen to the cheating information from malevolent nodes which guide to sink hole.
- Use of some protocol (cryptographic method with keys) that actually might try to confirm the superiority of the route with end-to-end acknowledgements enclosing latency information or reliability [6].
- It counts the list of suspected nodes, do a graph of Network flow and recognizing a sink attack by viewing missing data from an area attacked. This method is based on a CPU/ base station [3]. The base station/CPU overloads the network with a request note enclosing the IDs of the nodes which are affected. The affected nodes respond to base station with a note enclosing their IDs, ID of the next leap and the linked cost. Likewise the received information is then used from the base station to build flow graph of the network [7].
- Another approach is Received Signal Strength Indicator (RSSI) readings of messages. The projected solution requests collaboration of some Extra Monitor (EM) nodes rather than the ordinary nodes. It uses values of RSSI from 4 EM nodes to decide the situation of sensor nodes where the base Station is positioned at origin position (0, 0). This information is used as weight from the base station to detect this attack [7].
- Another scheme is a mobile agent which is proposed in [6]. It is a self controlling program segment. It navigates from node to node while transmitting or computation of data. Based on mobile agents a routing algorithm with manifold constraints is proposed. It uses mobile agents to collect information of all mobile sensor nodes to make every node responsive of the total network so that a valid node will not pay attention to the corrupt information from compromised node which heads to sinkhole attack. Using the Aglet mobile agent is developed which is proposed in [3]. Aglet, developed by IBM is a Java based system. Agents are called aglets here. The system suggests 2 algorithms. And they are: Agent navigation algorithm tells how a mobile agent provides information to nodes and visits every node and Data routing algorithm tells how a node utilize the information of the universal network to route data packets.

## 5.2. Hello Flood

**Attack:**

- Many WSNs routing protocols after deployment necessitate nodes to transmit HELLO packets, which is neighbor discovery based on radio range of the node.
- Laptop class attacker can transmit HELLO note to nodes and then promote high-quality path to sink.

**Counter Attacks:**

- This attack can be counteracted by using "identity verification protocol". It authorizes bi-directionality of a link with encrypted echo back mechanism even before taking consequential action based on a note received over that link [8].
- To guard against this, each request (REQ) message advanced by a node is encrypted with a code. In tree protocol when 2 sensor nodes share some general secrets then the original encryption key is spawned on the way. In this way, a node accessible neighbor can decrypt and confirm RREQ message while the foe will not be acquainted with the code and will be prohibited from initiation the attack. New key combined with the echo back mechanism can guard this attack [9].
- Route Error (RERR) message sends by the intermediate neighbor node and the source node reinitiates the path discovery process when there is an absence of hello packet during the periodical hello interval. In a haphazard manner the hello period values are altered and suggest this data to other nodes in network in a tenable manner. This categorizes and avoids the flooding attack [10].

### 5.3. Worm Hole

**Attack:**

- Scheme: tunnel packets established on one part of the system to another
- Located in a desired environment, wormhole can entirely mess routing
- These wormholes can utilize routing race circumstances which occurs when node takes routing decisions depending on the primary course advertisement
- Assailant may sway system topology by conveying routing data to the nodes before it would in fact get to them by multi hop routing
- This may persuade distant nodes that they are very close to the sink. This may also headed to sinkhole if node on the other end foes elevated quality way to sink
- Wormholes may induce 2 nodes that they are neighbors when they are far away from each other
- It can be used in conjunction with sybil attack
- Encryption alone cannot prevent this attack

**Counter Attacks:**

- To detect and for the solution of this attack, there is a easy exchange of 4 way handshaking messages. It is easy to implement and it does not entail any location information or time synchronization [9].
- Detection can be done on basis of data packet flow. In the proposed journal [11], by using network simulator NS2 whose code is written as tcl script this attack detection can be implemented. This attack is detected based on packet reception and dropped ratio, and also based on throughput. Number of packets is declining exponentially so WSNs can be protected by using this technique and this is how the attack can be prevented.
- In [12], a way out to this attack for adhoc networks where all nodes are ready with directional antennas.

- For perceiving and shielding against this attack Packet Leash [13] is one of the most suitable Method. This method has 2 types of leashes they are Geographic and Temporal. **Geographic Leashes:** All the nodes have a synchronized clock and knows it's exact location. Each node attaches its transmission time and existing position before transferring the packet. Receiving node figures out the path/ distance and also calculates the time for the packets to reach. To guess whether the predictable packet passed through wormhole or not, distance information can be used. **Temporal Leashes:** Here the correspondent puts the sending time with the packet and then the receiving node calculates the distance/ path of that packet by pretentious promulgation and exploiting the discrepancy between he time to send and receive the packet. This solution necessitates synchronization amongst nodes.
- Contrasting packet Leash, [14] presents SECTOR, it does not demand lock synchronization and information of the site. By implementing Mutual Authentication with Distance-Bounding (MAD) this attack can be prevented. By using the time of Flight and by using 1 bit challenge the prime node observes whether or not the second node is a neighbor and also calculates its distance to the second node.

## 5.4. Selective Forwarding/ Grey Hole

**Attack:**

- In WSNs multi hop model is widespread
- Threat: compromised node sends selected packets
- It is comprehend that nodes authentically send received messages
- Compromised node may decline to send packets, nevertheless neighbors might establish a different course

**Counter Attack:**

- this attack can be alleviated using multipath routing with amalgamation with random choice of trail to destination
- Additional solution is to check whether that the neighbor node sends the messages or not. The use of watchdog can prevent this attack
- According to [15], a distributed lightweight defense scheme in opposition to selective forwarding attack utilizes by the fellow nodes to supervise the broadcasting of the event packet and detect selective forwarding attack by monitoring packets forwarding of two nodes in the transmission path and send again these packets plunged by the attackers. Event packet is forwarded according to the routing calculated by the routing algorithm (OPA_uvwts) from source to destination. The transitional node is accountable for sending the event packet. Monitor node is accountable for the detection of probable selective forwarding attack and if this attack is identified, it resends event packet to the destination node and finally sends a disturbing message to its neighbor to notify the site of the foe and thus evade the attacker node in forwarding the incoming packets.
- In the same paper [15], they have mentioned chronological mesh test based selective forwarding attack detection scheme which is centralized to prevent this attack. Its works for cluster based sensor networks. These WSNs node sends the details of the packet drop through another pathway to cluster head if it doesn't see the send message from the next hop sensor node in an eternal interval. Cluster head runs this detection scheme of sequential mesh test against the apprehensive node just after receiving the drop report of the packet. Then sequential mesh test, according to the paper [17] instead of regulating in advance the total times of test extort a small samples to run the test. After testing the small sample it decides whether to continue the test or not. To detect selective forwarding attack WSNs nodes should listen to the network after forwarding the data packets. If the sender hasn't experiment the send message after a certain time then it can be suspected that the packet of the transitional relay node has been dropped. Then the sender node will report the event of the packet dropping to the cluster head through another course. According to [18], this scheme depends on the sequential mesh test which is hypothesis test and depending upon the ratio of packet drop the cluster head decides that the particular node launching selective forwarding attack or not. But the accurate ratio of the detection scheme becomes satisfactory when the package drop rate elevates the normal rate.

- In [19], a routing algorithm of AODV has been proposed to preserve the attack. The primary phase of this algorithm is "Counter threshold" and it utilizes the packet counter and the detection threshold to recognize the attacks. Second phase is "Query based" and it uses acknowledgment from the intermediate nodes to find out the attacker.

- When the presence of a spiteful node is detected by the algorithm then the primary node call upon the second phase- Query Based algorithm. Here the primary node will inquire the intermediate nodes about the value of the received packet. If all the intermediate nodes are queried together then it will augment the overhead of the algorithm. So to give better performance, counter frequency should be used to select the intermediate nodes. And thus it will be easier to select the attacker.

### 5.5. Acknowledgement Spoofing

**Attack:**

- Link layer acknowledgments is used by the routing protocols
- Objective: it is convinced that whether a fragile link is strong or a quiet node is alive or not
- A foe may spoof ACKs (acknowledgement)
- Accordingly fragile link may also be selected for routing
- Packets forwarded trough that link may be lost/ corrupted.

**Counter Attacks:**

- This attack can be prohibited via good encryption and authentication techniques for communication. Since the base stations are reliable so a foe may not be able to spoof broadcast/ flooded messages from any base station. This necessitates asymmetry wherein no node should be able to spoof any message from base station; at the same time all the nodes should be able to authenticate them. Genuine broadcast is functional for the interactions of restricted node. This attack occurs on link layer of WSNs. This attack is possible on hierarchical routing protocol; location based and QoS aware routing protocols, Network flow.

- Based on [20], SNEP, TESLA, Random key distribution are the security schemes.

## 5.6. Sybil

**Attack:**

- Scheme: a single node acts to be present in another part of the network.
- It affects geographical routing protocols mostly.

**Counter Attacks:**

- Trusted certification has the probability to eliminate Sybil attacks [21]. It is cited as the common clarification. On the other hand it depends on a centralized authority that makes sure each body is assigned to exactly one identity, as specified by owning a certificate. In [21], researchers present no other method of guarantying such uniqueness and it is performed by manual process. This create a performance bottleneck and costly for a large-scale systems. Furthermore to be more efficient the endorsed authority make sure the stolen/ lost identities are exposed and revoked. This approach can eliminate the attack easily when the performance and security implications are solved.
- Authors of [22] and [23] proposed testing of IP addresses in different autonomous systems. It necessitates heterogeneous IP addresses stops some particular attacks however does not dampen other zombie networks and also it put a restriction in the usability of the application.
- A validation technique is also used to avert this attack and also discharge masquerading hostile bodies. A local unit may recognize a distant identity based on a fundamental authority ensuring a one to one correspondence between a unit and a body and may also offer reverse lookup. Individuality may be legalized indirectly/ directly. In direct validation: the local unit enquires the fundamental authority to legalize the isolated identities. In indirect validation the restricted unit relies on previously conventional identities which in turn guarantee the validity of the distant individuality in question.
- The authors of [11] and [12] proposed PKI (Public Key Cryptography/ Infrastructure) to defend against this attack. Various algorithms are proposed by them. In this one central authority is responsible for giving certificates to each vehicle. Certificate contains PKI, a set of physical attributes of a vehicle. Vehicular PKI is very heavy to deploy due to the existence of large number of vehicles by different manufactures and countries.

## 5.7. Jamming

**Attack:**

- Scheme: attacker tries to broadcast signal to a base station at the same frequency sub band/ band as the transmitter
- It causes radio interference in the network. It disrupts the radio communication.
- This attack is used by a laptop which grasps higher energy to upset incessantly the network. Also it can be done with a simple node sourcing a partial damage which can be also deadly to the WSNs (like random distributed jammed node).

- In [24], author presented diverse jamming strategies: constant jamming by emanating constantly a radio signal, misleading jamming: by inserting normal packets to the conduit without any gap between them, random jamming: where the foe exchanges to save power consumption between sleeping and jamming and lastly reactive jamming: which will send only when it senses activity of the channel and will stay silent when the channel is inactive.

**Counter Attacks:**

- So many solutions were proposed to protect against these attacks. Typical resistance involves variations of spread-spectrum communication: frequency-hopping spread spectrum (FHSS) which includes forwarding data by switching rapidly a carrier sense amongst a lot of frequency channels/ code spreading. This technique used in military applications due to the complexity and high cost, example MICA2 mote is the only sensor which alternates efficiently between 2 frequencies and for every additional frequency will require extra processing.
- According to the paper [24], Nodes can try to map out the jammed area by separating the contaminated region. Such a protocol was also presented in paper [25].
- Channel surfing method is motivated by the frequency hopping modulation which is a solution for this attack. The difference with FHSS is that it does not involve a repeated transformation of the carrier sense and it functions at the link layer.

### 5.8. Tampering

**Attack:**

- Scheme: Foe can control these motes and try to find out perceptive information: secret key shared between nodes.
- It can be classified in 2 classes: invasive attacks, which entail access to the hardware apparatus like chips and which require high tech and exclusive apparatus used in semiconductor industry, and non-invasive which requires less time and more flexible.
- Attack via JTAC1: testing access port (TAP) which enable a foe to control over the sensor node.
- Other attacks via exploiting the Bootstrap Loader (BSL): enables READ and WRITE on the micro controller's memory. Foe can also bother the external flash/ EEPROM where precious data are stored.
- Causes the entire scratch to the network services.

**Counter Attacks:**

- A straightforward way to comprehend this attack is to eavesdrop on the conductor wire linking between the external memory chips and the micro controller.
- The authors of [26] presented a new key management protocol detecting the inoculation of spiteful nodes in the network. There is no comprehensive solution against these attacks. Typical precautions are applied: disabling the JTAG interface or using high-quality password for the bootstrap loader.

## 5.9. Collisions

**Attack:**

- Scheme: A foe forwards its own signal when it listen a legitimate node transmitting a message to make interferences. In theory, causing collisions to 1 byte is sufficient to create CRC error which cripples the message.
- It is similar to the continuous channel attack. A collision is when two nodes effort to broadcast on the same frequency and at the same data rate.
- Due to collision change will occur in the portion of the data so that disparity error occurred. The data packet will then be superfluous as invalid.

**Counter Attacks:**

- All countermeasures used against jamming attacks can be applied to these attacks.
- Error correcting codes is another solution [27] for this attack which is well structured in state of affairs where errors occur on an inadequate number of bytes but this solution is an expensive one which presents communication overheard and additional processing.

## 5.10. Exhaustion

**Attack:**

- Scheme: According to [28], this attack consists in introducing collisions in frames towards the end of transmission and force the sensor node to retransmit continuously the packets until it is death. This exhaustion attack is launched using an ordinary sensor node or by a laptop.
- This type of attack leads to starvation by continuously sending data or request over the channel.

**Counter Attack:**

- Limit the MAC admission control rate so the sensor network ignores excessive requests from the adversary and prevent energy loss.
- Allow a small slot of time for each sensor node to access to the channel and broadcast data, so it confines the possibility of long use of MAC channel.
- This attack has been solved by two other techniques: Rate Limiting and Time Division Multiplexing [29]. Rate Limiting by MAC admission controls excessive requests and repeated transmission.

# 6. A COMPREHENSIVE STUDY OF ATTACKS

The above presented attacks are summarized in table 1, including the attack type, the corresponding protocol layer, its characteristics, a possible solution and the damage it may cause.

Table 1.  A summary of WSN attacks.

| Attack | Layer | Features/ Characteristics | Solution | Result of the attack |
|---|---|---|---|---|
| **Sybil** | Routing layer | Malevolent node supposes numerous individualities. The aim is to fill the memory of the neighbouring node with useless data | Validation techniques is the solution of Sybil attacks and it dismisses masquerading antagonistic entities | The result of this attack can be biased or complete degradation of the network's service, depending on the site of the commencement of the attack. It can overcome the redundancy method of the distributed storage systems |
| **Sinkhole (Blackhole)** | Routing layer | Malevolent node acts as a black hole and tries to draw all probable traffic through a compromised node creating a metaphorical sinkhole with the adversary at the centre | Use a quintessential scheme which will make the entire node aware of the entire network so that the corrupt information from malicious node will not listen to the valid node. To confirm the quality of route with end-to-end acknowledgements some protocol (cryptographic method with keys) can be used which contains reliability or latency information | Creates a blackhole/sphere of influence in the sensor network |
| **Hello flood attack** | Routing layer | It uses HELLO packets as a warhead to persuade the WSNs sensors. Protocols depending on the localized information swaps between adjacent nodes for topology upholding or flow control are also area under discussion. It can also be thought of as one-way, broadcast wormholes. | Hello flood attack can be counteracted using "identity verification protocol". This substantiates bi-directionality of the linkage with encrypted echo-back mechanism, prior to captivating meaningful action based on a message received over that connection. | Every node thinks that the attacker is within one-hop radio communication range. If the attacker subsequently advertises low-cost routes, nodes will attempt to forward their messages to the attacker. |
| **Worm hole attack** | Routing layer | Malicious nodes eavesdrop the packet and can tunnel messages received in one part of the network over a low latency link and retransmit them in a different part. It can be used to exploit routing race conditions. | To detect and for the solution of this attack, there is a simple four-way handshaking messages exchange. Use of private channel can be another solution of Worm hole attack. | Routing race conditions characteristically occur when a node takes some action based on the first example of a message it receives and afterwards ignores later instances. The goal of this attack is to challenge cryptography protection and to confuse sensor's network protocols. |
| **Selective Forwarding attack** | Routing layer | Malicious nodes forwards most messages and selectively drops, which means throwing away some of the data | To select different paths randomly toward destination, we can use multipath routing .The probability of message that will encounter an adversary along all routes decreases by this one. Another solution is we can use by forwarding message toward neighbors that can be done by the monitor nodes. Watchdog can be used as a supervisor of the system. | The result will be a total broken of all service offered by the network |
| **Acknowledgement spoofing** | Routing layer | Due to the inherent broadcast medium, an adversary can spoof link layer acknowledgments for ''overheard'' packets addressed to neighbouring recursively with each node marking its parent as the first node from which it hears a routing update. | The most obvious solution to this problem would be authentication via encryption of all sent packets | It creates routing loops, attract or repel network traffic, extend or shorten source routes, generate false error messages, partition the network, increase end-to-end latency |
| **Jamming** | Physical layer | Malicious node tries to transmit signals to the receiver at the same frequency band or sub band as the transmitter uses and causes interference | To mitigate Jamming evolutionary algorithm, the ant system, Symmetric encryption algorithm, Brute force attacks against block encryption algorithms is used. | The adversary must be capable of classifying transmitted packets in real time, and corrupting them before the end of their transmission. The increased noise floor results in a flattered signal to noise ratio, which will be indicated at the client |

| Device Tampering | Physical layer | The simplest way to attack is to damage or modify sensors physically and thus stop or alter their services. Base stations or aggregation can be attacked as well | Use Tamper-resistant devices, Tamper proof systems | Stop the service of the sensors |
|---|---|---|---|---|
| Collisions attack | Link layer | Message transmission by two nodes on a same frequency simultaneously | Use of error correcting codes | It changes packet's fields, and also alters the acknowledgement message |
| Exhaustion attack | Link layer | Continuous retransmission and repeated collisions until the sensor node becomes dead | Rate Limiting and Time Division Multiplexing | Continuously retransmission, message modification; acknowledgement message corruption/change |

# 7. CONCLUSIONS

In the case of WSN, security has been an increasingly significant subject. Due to resource limitations, it is quite impossible to provide a strong security to a WSN. In the present paper, attacks and defences referenced from 1997 till 2013 have been summarized and particular solutions were proposed. As all WSNs attacks and counter attacks have been presented the extensive study could offer a review of the relevant topic for future research on WSN security. Future work should focus on finding a solution for combinational link layer attacks, designing the MAC36 protocol, or securing WSNs links against collision and DoS attacks. More comprehensive research is also necessary to measure the efficiency of algorithms in terms of resources available.